\documentclass[a4paper,aps,pre,reprint,twocolumn,raggedfooter,showpacs]{revtex4-1}
\usepackage{graphicx}
\usepackage[caption=false]{subfig}
\usepackage{latexsym}
\usepackage{amsmath}
\usepackage[makeroom]{cancel}
\usepackage{relsize}
\usepackage{letltxmacro}
\usepackage{amssymb}

\usepackage[usenames]{color}

\LetLtxMacro{\oldsqrt}{\sqrt} 
\renewcommand{\sqrt}[1][\ ]{%
  \def\DHLindex{#1}\mathpalette\DHLhksqrt}
\def\DHLhksqrt#1#2{%
  \setbox0=\hbox{$#1\oldsqrt[\DHLindex]{#2\,}$}\dimen0=\ht0
  \advance\dimen0-0.2\ht0
  \setbox2=\hbox{\vrule height\ht0 depth -\dimen0}%
  {\box0\lower0.71pt\box2}}

\begin{document}

\title{Elasticity of fibrous networks under axial prestress}
\author{M.\ Vahabi}\affiliation{Department of Physics and Astronomy, Vrije Universiteit, Amsterdam, The Netherlands}
\author{A.\ Sharma}\affiliation{Department of Physics and Astronomy, Vrije Universiteit, Amsterdam, The Netherlands}
\author{A.\ J.\ Licup}\affiliation{Department of Physics and Astronomy, Vrije Universiteit, Amsterdam, The Netherlands}
\author{A.\ S.\ G.\ van Oosten}\affiliation{Institute for Medicine and Engineering, University of Pennsylvania, Philadelphia, PA, United States.}
\author{P.\ A.\ Galie}\affiliation{Institute for Medicine and Engineering, University of Pennsylvania, Philadelphia, PA, United States.}
\author{P.\  A.\ Janmey}\affiliation{Institute for Medicine and Engineering, University of Pennsylvania, Philadelphia, PA, United States.}
\author{F.\ C.\ MacKintosh}\affiliation{Department of Physics and Astronomy, Vrije Universiteit,\;\!\!\! Amsterdam, The Netherlands}
\date{\today}

\begin{abstract}
We present theoretical and experimental studies of the elastic response of fibrous networks subjected to uniaxial strain. Uniaxial compression or extension is applied to extracellular networks of fibrin and collagen using a shear rheometer with free water in/outflow. Both axial stress and the network shear modulus are measured. Prior work [van Oosten \textit{et al., Scientific Reports}, 2015, \textbf{6}, 19270] has shown softening/stiffening of these networks under compression/extension, together with a nonlinear response to shear, but the origin of such behaviour remains poorly understood. Here, we study how uniaxial strain influences the nonlinear mechanics of fibrous networks. Using a computational network model with bendable and stretchable fibres, we show that the softening/stiffening behaviour can be understood for fixed lateral boundaries in 2D and 3D networks with comparable average connectivities to the experimental extracellular networks. Moreover, we show that the onset of stiffening depends strongly on the imposed uniaxial strain. Our study highlights the importance of both axial strain and boundary conditions in determining the mechanical response of hydrogels.
\end{abstract}

\maketitle

\section{Introduction}
Proteins, essential molecules of living organisms, can be found in the form of fibrous networks both inside and outside the cells\cite{alberts2007molecular,bausch2006bottom,kasza2007cell,chaudhuri2007reversible,lieleg2010structure,fletcher2010cell,frantz2010extracellular}. The cytoskeleton, blood clots and extracellular matrices of tissues all consist of fibrous protein networks. The mechanics of these systems depend not only on the elastic properties of the individual fibres but also on geometrical properties of the network such as average connectivity, cross-linking and branching distance\cite{gardel2004elastic,chaudhuri2007reversible,tharmann2007viscoelasticity,broedersz2011criticality,gardel2006prestressed,koh2012branching,wagner2006cytoskeletal,lin2010origins,hatami2009effect,erk2010strain}. One of the generic mechanical features of these structures is their nonlinear strain-stiffening behaviour. The nonlinear stiffening is ubiquitous in biological systems and is apparent in the rapid increase of the material stiffness when subject to strain\cite{fung1967elasticity,janmey1994mechanical,shah1997strain,sacks2000biaxial,gardel2004elastic,roeder2002tensile,storm2005nonlinear,motte2013strain}. This makes them compliant to small deformations and resistant to large deformations. An increased resistance to large strains can act as a protection mechanism against tissue damage\cite{storm2005nonlinear,koh2012branching}. The property of a strain dependent stiffness of biopolymers has also inspired recent efforts to create synthetic polymers with similar properties\cite{kouwer2013responsive,jaspers2014ultra}. 

Several experimental and theoretical studies have focused on understanding the nonlinear mechanics of filamentous networks\cite{broedersz2014modeling,wen2012non,kang2009nonlinear,storm2005nonlinear,gardel2004elastic,conti2009cross,kabla2007nonlinear,onck2005alternative,Licup04082015,sharma2016strain,lin2011control,pritchard2014mechanics}. 
This striking behaviour can be understood for both thermal (entropic) and athermal (enthalpic) models. Affine thermal models are based on the nonlinear force-extension relation for individual semiflexible filaments between network junctions\cite{mackintosh1995elasticity,morse1998viscoelasticity}, where the entropic stretching of the filaments or cross-linker proteins leads to a reduction in the amplitude of the transverse thermal undulations that in turn gives rise to a dramatic entropic strain stiffening\cite{gardel2004elastic,storm2005nonlinear,lin2010origins,broedersz2014modeling}. The origins of stiffening in athermal models in contrast lie in non-affine collective deformation of the fibrous networks composed of interconnected elastic rods\cite{head2003deformation,wilhelm2003elasticity,das2007effective,heussinger2007role,conti2009cross,mao2013elasticity,mao2013effective}, which can result in a nonlinear mechanical response at the network level, even for networks composed of purely linear elastic elements\cite{onck2005alternative,broedersz2014modeling,heussinger2007nonaffine,conti2009cross}. 

Many experiments have been performed on reconstituted networks of biopolymers\cite{janmey1994mechanical,storm2005nonlinear,gardel2004elastic,chaudhuri2007reversible,wagner2006cytoskeletal,lin2010origins,gardel2006prestressed}. These constitute a new class of biological soft matter systems with remarkable material properties. Moreover, studies on these yield valuable input and tests for theoretical modeling of extracellular matrices and the cytoskeleton \emph{in vivo}\cite{broedersz2014modeling,hatami2011cytoskeletal,tharmann2007viscoelasticity,Licup04082015,sharma2016strain,motte2013strain}.

Although intracellular networks, extracellular matrices and whole tissues show many similar mechanical properties, there are several important differences. Extracellular matrices tend to be much more open structures with larger pore size, making them more compressible than the typically finer intracellular meshworks on the same time-scales: as the incompressible fluid flows in and out, the networks can effectively change their volume. Also, extracellular biopolymers tend to have a larger persistence length. This is in particular true of collagen, which forms networks that can be treated as athermal and fully mechanical\cite{motte2013strain,stein2011micromechanics,Licup04082015,sharma2016strain,van2016uncoupling}. In several studies, the mechanical response of tissues under compression/extension has been investigated\cite{comley2012compressive,fung1967elasticity,bonfield1977anisotropy,elliott2001anisotropic,soden1974tensile,lally2004elastic,mihai2015comparison,perepelyuk2016normal,pogoda2014compression}. 
It has been found that tissues exhibit stiffening under compression\cite{comley2012compressive,fung1967elasticity,mihai2015comparison,perepelyuk2016normal,pogoda2014compression}. In addition, some reports have also reported some stiffening under extension\cite{comley2012compressive,fung1967elasticity,lally2004elastic}. Incompressible continuum models and finite element methods have been exploited to describe such behaviour\cite{comley2012compressive,wu1976mechanical,shergold2006uniaxial,ogden1972large}. 

In contrast, biopolymer networks, including collagen matrices similar to the networks in whole tissues soften under compression and stiffen under extension\cite{van2016uncoupling} on time scales long enough for influx/efflux of interstitial fluid. As we discuss here, one difference between the two systems, tissues and extracellular networks, is closely related to the difference in the applied boundary condition. This behaviour is also completely different from that of the linear synthetic polymers such as polyacrylamide that do not show any stiffening for the same range of axial strains\cite{van2016uncoupling}. Here, we consider disordered lattice-based network models with comparable average connectivities (coordination number) to real biopolymer networks\cite{lindstrom2010biopolymer}. The networks consist of bendable and stretchable fibres. By applying a fixed (\textit{lateral}) boundary condition on our network under axial strain, we can account for the mechanical behaviour we observe for reconstituted fibrin and collagen hydrogels. We also show that applying a global volume constraint on the network results in stiffening for both compression and extension.

Here, we focus on the effect of prestress, in the form of network extension and compression, on the properties of these networks. These are important aspects that have not received as much attention as shear rheology in recent work\cite{van2016uncoupling}. Prestress, in which residual stresses exist in an unloaded sample, happens frequently in cells and tissues\cite{robertson2013multiscale,fung1977first}. Prestress can be natural and useful, e.g., in the cardiovascular system, where it can increase pressure resistance\cite{destrade2012uniform}, and in cells, where myosin II motors can increase cell/gel stiffness by the generation of active, internal stresses\cite{jansen2013cells,kollmannsberger2011nonlinear,mizuno2007nonequilibrium}. Moreover, in blood clots, active, contractile stresses due to platelets are vital for wound closure\cite{jen1982structural}. But, prestress alteration can also be harmful as in pathologic conditions including hypertension and atherosclerosis\cite{liu1989relationship,hong1997altered}. Generally, the mechanical properties of prestressed systems differ from relaxed systems, due to intrinsic nonlinearities. Experiments have shown that biopolymer networks polymerized in the rheometer develop normal stresses\cite{van2016uncoupling}. Though the origins of such normal stress are not always understood, it is evident that the mechanical response of the network is influenced by such initial stresses\cite{Licup04082015}. It has also been shown that active agents such a molecular motors acting on networks can give rise to increased stiffness\cite{mizuno2007nonequilibrium,mackintosh2008nonequilibrium} and normal stress\cite{bendix2008quantitative}. However, even in the absence of active agents, normal stresses can arise as in the polymerization process or simply due to an initial extension or compression applied to the network prior to subjecting the network to a shear deformation. The latter is the approach followed by Ref.\cite{van2016uncoupling}. Here, we systematically investigate the nonlinear mechanics of networks that have been subjected to an initial uniaxial deformation. We show that the prestress due to the initial deformation impacts the linear shear modulus and the onset of stiffening in the nonlinear shear stiffening curves.

The paper is organised as follows: Section \ref{Model} briefly describes the network model used here. Section \ref{Exp} concisely explains the experimental methodology. In Section \ref{Results}, we show the effects of prestress in the form of extension and compression on linear shear modulus and the nonlinear strain-stiffening both in experiments and simulation. We conclude in Section \ref{Conclusions} with a brief summary and present our conclusions.

\section{Model}\label{Model}
To study the mechanical properties of biopolymer networks, we employ a minimal model to generate disordered, lattice-based networks. The networks are based on 2D triangular or 3D face-centred cubic (FCC) lattices with lattice spacing, $l_0$. Starting from these networks, we use a phantomization process to generate phantom networks with a local coordination number or connectivity $z=4$, since higher connectivities are unphysical in experimental systems for networks consisting of cross-linked fibres ($z=4$) and branching ($z=3$). This phantomization is done for 2D networks by modifying triangular lattices such that at every lattice vertex, where three fibres cross, one filament is chosen at random and disconnected from the other two, allowing it to move freely as a phantom chain with no direct mechanical interaction with other two filaments\cite{broedersz2011molecular}. For the 3D FCC lattice, where six fibres cross at each node, we randomly choose three independent pairs of cross-linked filaments\cite{broedersz2012filament}. The phantomization procedure, sets the connectivity of the respective networks precisely to $z=4$. We further dilute the networks by random removal of bonds (fibre segments) between vertices. This is done to achieve the desired connectivity $z$, where $3<z<4$. It is also possible to reach this connectivity by only using the dilution process. The resulting networks are then called 2D diluted triangular lattices. We also ensure that there are no fibres spanning the full network\cite{broedersz2012filament}. 

Importantly, this procedure results in a disordered network structure, in spite of the initial, regular lattice structure. Moreover, the resulting connectivity lies below the threshold of marginal stability for purely pairwise central-force interactions in both 2D and 3D, as identified by Maxwell\cite{maxwell1864calculation,thorpe1983continuous}: this threshold is twice the dimensionality of the system. This means if spring-like interactions were the only relevant contribution to the Hamiltonian, these structures should be floppy and unstable. Networks of biopolymers are, however, stable 3D structures with average connectivity below this isostatic threshold\cite{lindstrom2010biopolymer}. In fact, such sub-isostatic networks can be rigid due to additional stabilising interactions, such as bending\cite{wilhelm2003elasticity,head2003deformation,broedersz2011criticality}, internal or applied stresses\cite{sheinman2012actively,Licup04082015,alexander1998amorphous} or thermal fluctuations\cite{DennisonPhysRevLett}. Here, we include bending interactions in our model. To reduce any edge effects, periodic boundaries are imposed with Lees-Edwards boundary conditions\cite{lees1972computer}. The cross-links or branching points are permanent in our networks and they hinge freely with no resistance. Prior simulations of networks consisting of cross-linked or branched fibres\cite{broedersz2011criticality,das2012redundancy,rens2016nonlinear}, with and without freely-hinged cross-links, has shown very similar behaviour in mechanical properties for the same average connectivity $z$. This suggests that additional bending interactions at the cross-links, as can be expected for both fibrin and collagen, will not significantly affect our model predictions. 
\begin{figure}[t]
\centering
\includegraphics[width=0.4\textwidth]{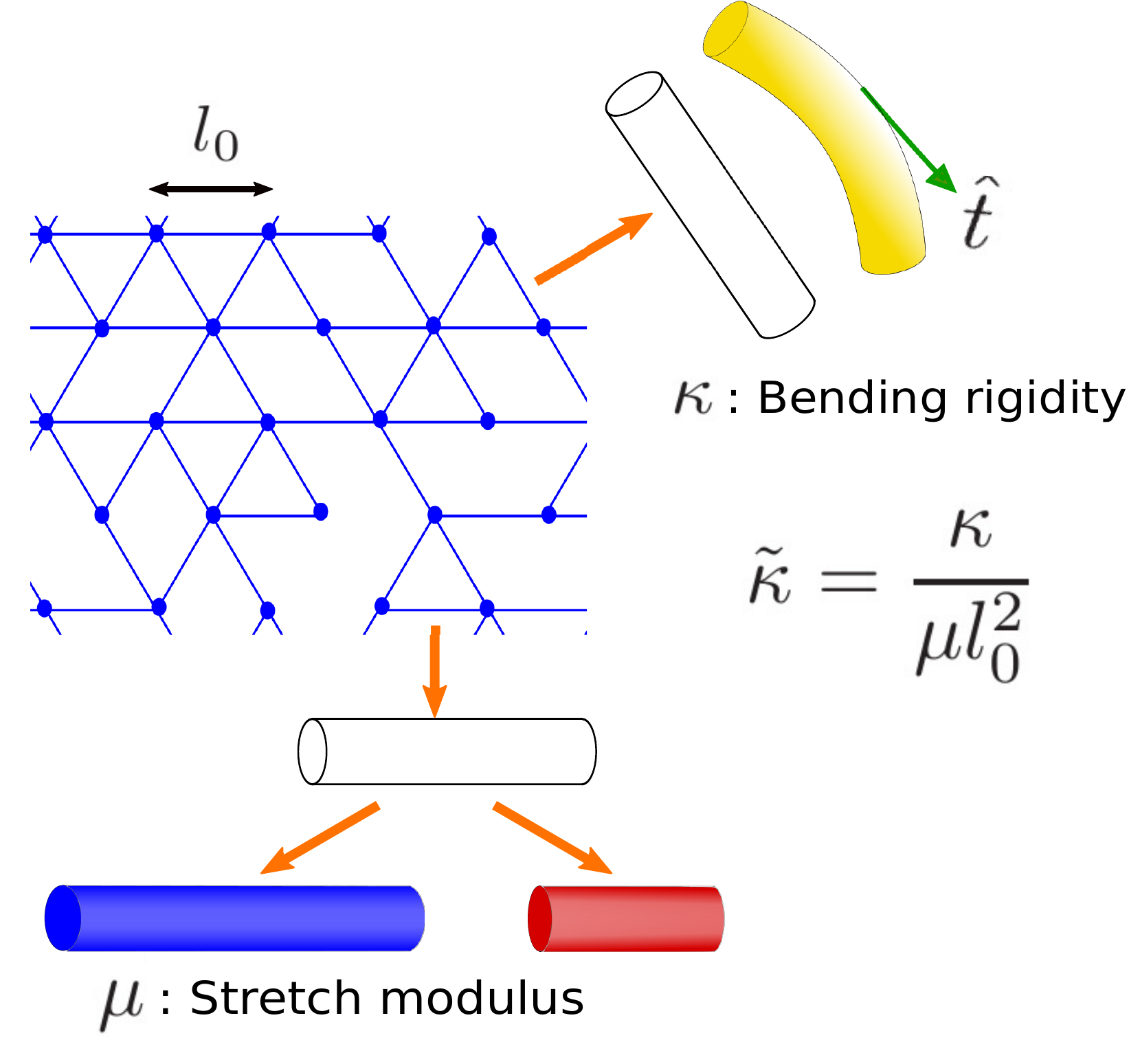}
\caption{(colour online) 2D Schematic representation of the model: The network is a 2D diluted triangular lattice with lattice spacing, $l_0$. Each bond is assumed to act like a spring with stretch modulus, $\mu$. The blue and red colouring of the filament shows its extension and compression. fibres can also bend at the hinges with bending rigidity, $\kappa$ which is shown by the yellow bent fibre. The dimensionless parameter, fibre rigidity, $\tilde{\kappa}$ is then defined as the ratio between the bending rigidity, $\kappa$ and stretch modulus, $\mu$ where $l_0^2$ is used for dimensional purposes, $\tilde{\kappa}=\kappa/(\mu l_0^2)$.}
\label{fig:schematic}
\end{figure}%

The filaments in the network are described by an extensible wormlike chain (EWLC) model (bending and stretching contributions) and the Hamiltonian of the system $\mathcal{H}$ is obtained by summing over all the fibres\cite{head2003deformation}, $f$:
\begin{equation}
\label{eqn:Efibre}
\mathcal{H}=\sum_{f}\left[\int \frac{\kappa}{2} \left| \frac{d\hat{t}}{ds_{f}} \right|^{2} ds_{f} + \int \frac{\mu}{2} \left(\frac{dl}{ds_{f}} \right)^{2} ds_{f}\right].
\end{equation}
Here, $\kappa$ is the bending rigidity of the individual filaments, $\mu$ is their stretch modulus and, $\hat{t}$ and $dl/ds_f$ are the unit tangent and longitudinal strain respectively at a point $s_{f}$ along the fibre contour. Here, we only consider athermal networks, which have been shown to successfully capture the mechanics of collagen networks\cite{Licup04082015,sharma2016strain}. Although the individual elements of the model are linear, the network mechanics are highly nonlinear. The schematic of the model in $2$D is shown in Fig.\ref{fig:schematic}. 

The dimensionless bending rigidity is defined as $\tilde{\kappa} = \kappa / (\mu l_{0}^{2})$ which we vary in our networks while keeping $\mu=1$ constant. For 3D networks based on FCC lattices, $l_0$  is the same as the distance between the cross-links, $l_c$. For 2D phantom networks, the average distance between the cross-links is somewhat larger\cite{licup2016elastic}, $l_c\simeq 1.4 l_0$ for $z\simeq 3.3$. Given a homogeneous cylindrical rod of radius $r$ and Young's modulus $E$, from classical beam theory\cite{landau1970course}, $\mu=\pi r^2 E$ and $\kappa=\frac{\pi}{4} r^4 E$. From this, $\tilde{\kappa} = \frac{1}{4}r^2/ l_{0}^{2}$, which is proportional to the protein volume fraction $\phi$, which can be seen as follows. Within a volume $l_0^3$ in the network, there will be of order one fibre segment of volume, $\pi r^2 l_0$, corresponding to a volume fraction\cite{van2016uncoupling} $\phi \sim r^2/l_0^2\sim \tilde{\kappa}$. Hence, the most relevant values of $\tilde{\kappa}$ for biopolymer systems range from\cite{Licup04082015,broedersz2011molecular} $10^{-4}$ to $10^{-2}$.  

To find the elastic stresses or responses of these networks, the relevant deformation is applied to the network and then the energy is minimised using the conjugate gradient minimisation method\cite{press1996numerical}. We are in particular interested in shear stress, $\sigma_s$, the storage modulus, $G$, and normal stress, $\sigma_N$, all of which are obtained using the minimised total elastic energy, $E$. Shear stress is calculated from the derivative of the minimised elastic energy density $E/V$ of the network, where $V$ is the system area (volume) in 2D (3D), with respect to the applied shear strain, $\gamma$:
\begin{equation}
\label{eqn:sigmas}
\sigma_s=\frac{1}{V}\frac{\partial E}{\partial \gamma}.
\end{equation}
From this, the storage modulus is obtained as the ratio  
\begin{equation}
\label{eqn:G}
G=\frac{\sigma_s}{\gamma}.
\end{equation}
The normal stress, $\sigma_N$, is calculated from the derivative of the energy with axial strain $\varepsilon$:
\begin{equation}
\label{eqn:sigman}
\sigma_N=\frac{1}{V}\frac{\partial E}{\partial \varepsilon}.
\end{equation}

Our simulation results are in units of $\mu l_0^{1-d}$, where $d$ is the dimension of the system. All the simulation results are carried out on large enough systems sizes to minimise finite size effects. In an undiluted network, we use in 3D, $30^3$ nodes and $50^2$ nodes in 2D for all the reported results unless otherwise specified. The probability of existing bonds in 3D networks is $p=0.85$, in 2D phantom networks is $p=0.9$ and in 2D diluted triangular networks is $p=0.58$.

In the simulation, both 2D and 3D networks are studied, although biopolymer networks are inherently 3D structures. Recent computational studies of the lattice-based networks have shown remarkable quantitative agreement between 2D and 3D networks, both in their linear and nonlinear mechanical behaviour, provided that the networks have the same connectivity $z$ and are below the 2D isostatic threshold\cite{sharma2016strain,licup2016elastic}. The elasticity of 2D networks can be mapped to those from 3D by correctly accounting for the line density\cite{licup2016elastic} $\rho\propto l_0^{1-d}$. Moreover, the overall elastic properties of lattice- and off lattice-based network models have been demonstrated to be similar\cite{licup2016elastic,Licup04082015,sharma2016strain}.

For comparison with experiments, we use fixed lateral boundaries, for which the ratio of normal stress to axial strain gives the \emph{longitudinal} modulus $M=\sigma_N/\varepsilon$ for small strains $\varepsilon$. These boundary conditions are most relevant to extracellular networks of collagen and fibrin in a rheometer, for which the lateral dimension is typically much larger than the axial dimension (i.e., gap size). For networks that adhere to the axial boundaries (rheometer plates), this aspect ratio, together with the open network structure and flow of fluid in and out during rheological measurements, leads to effectively fixed lateral boundaries and a vanishing of the (apparent) Poisson ratio \cite{van2016uncoupling}. Here, we investigate the effect of uniaxial deformation on the shear and normal stresses and storage modulus of the networks. To apply uniaxial deformation to our networks, the length of the system in the direction perpendicular to the shear stress is initially rescaled. For example, to have a system subject to $10\%$ compression (extension), the axial length of the system is decreased (increased) accordingly. After applying this global deformation, the energy of the network is minimised before applying any shear measurements. This is similar to having a system in the prestressed state before these measurements. Then the effect of this prestress on the linear shear modulus and nonlinear shear strain-stiffening curves are investigated.

\section{Experimental}\label{Exp}
The experimental data were acquired similarly as in Ref.\cite{van2016uncoupling}. Briefly, a strain-controlled rheometer (RFS3, TA Instruments, New castle, DE, USA) was used in a parallel-plate configuration with plate diameters of 8 mm, 25 mm or 50 mm. Uniaxial strain was applied by changing the gap between the plates after sample polymerization, shear strain was applied by rotating the bottom plate. The upper plate was connected to a force sensor measuring both torque and normal force. 
To prepare collagen networks, collagen type 1 (isolated from calf skin, MP Biomedicals, Santa Ana, CA, USA), 10X PBS, 0.1M NaOH and ddH2O were warmed to room temperature and added in appropriate ratios to yield a 2.5 mg/ml collagen concentration in 1X PBS solution with a neutral pH. To prepare fibrin networks, fibrinogen stock solution (isolated from human plasma and plasminogen depleted, CalBioChem, EMD Millipore, Billerica, MA, USA,) 1X T7 buffer (50mM Tris, 150 mM NaCl at pH 7.4), CaCl2 stock and thrombin (isolated from salmon plasma  SeaRun Holdings, Freeport, ME, USA) were added at appropriate ratios to yield 2.5 or 10 mg/ml fibrinogen, 30mM Ca2+ and 0.5 U thrombin per mg of Fbg. The measurements are done in the shear rheometer or a tensile tester (Instron, Inc.) and the samples are completely surrounded by the buffer thus the fluid can flow in and out freely. By applying compression or extension by changing the gap size between the rheometer plates, axial strain is imposed. 

\section{Results and Discussion: elastic properties of networks under extension and compression}\label{Results}
Under typical physiological conditions, tissues in the body are constantly subjected to complex deformations. It is thus important to see how the mechanical properties of such systems vary under application of both shear as well as axial strain. As mentioned in the introduction, reconstituted networks of biopolymers are good candidates for studying real biopolymer scaffolds. In this work, we investigate the role of prestress generated by the axial strain on the mechanical properties of extracellular networks of collagen and fibrin. 

We compare experimental results with simulation results from 2D and 3D networks (see Model section). In order to compare the results from simulation to our experiments, we apply fixed lateral boundary conditions. Our 2D and 3D networks have comparable average connectivities as those observed in the extracellular networks\cite{lindstrom2010biopolymer}. Both experimental and computational results show softening for compression and stiffening for extension (which is different from the results obtained from the measurements on tissue\cite{comley2012compressive,fung1967elasticity,lally2004elastic,mihai2015comparison,perepelyuk2016normal,pogoda2014compression}. We focus on the impact of prestress generated due to the uniaxial strain on the mechanical properties of these systems. In subsection \ref{A} the impact on the linear shear modulus and in \ref{B} the strain-stiffening curves are considered. In both subsections (\ref{A} and \ref{B}), the effect of axial strain on normal stresses is also investigated. In subsection \ref{C}, we discuss the effect of global volume boundary condition on the simulation results which imposes the incompressibility condition on the networks.

\subsection{Effect of prestress on linear shear modulus}\label{A}
We investigate the role of uniaxial strain in the form of extension or compression on the linear shear modulus. Experimentally, the samples are subjected to an incremental series of compression/extension. At any given axial deformation, the dynamic shear moduli is measured at an amplitude of $2\%$ shear strain and frequency of $10 \enskip \mathrm{rad/s}$. In between each compression/extension step, the networks are allowed to relax for 100 to 1200 seconds depending on the step size or sample. A tensile tester is used to measure the axial stress of the samples after relaxation for similar levels of axial strain. 
\begin{figure}[htbp]
\includegraphics[width=.4\textwidth]{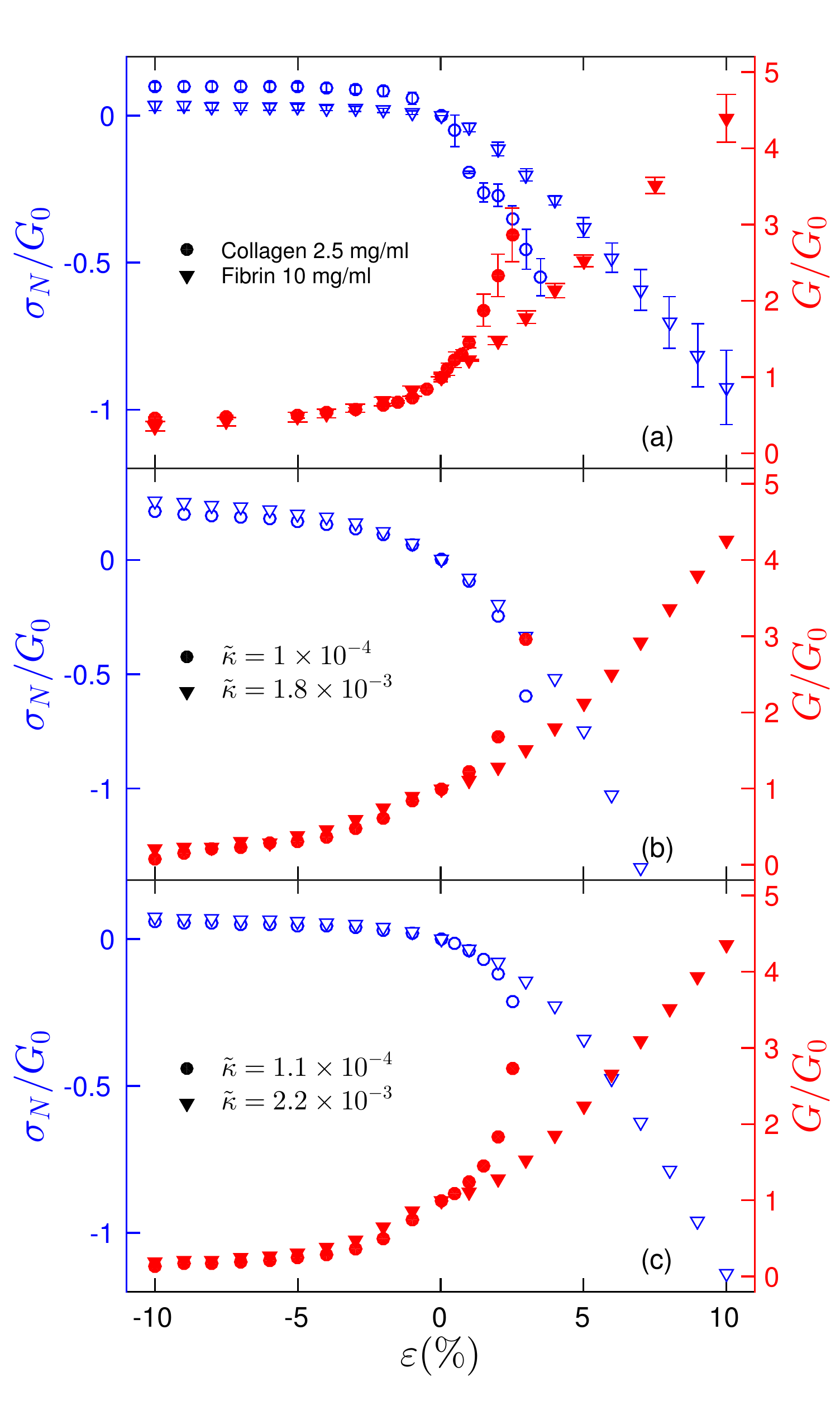} 
\caption{(colour online) Normal stress ($\sigma_{N}$) (unfilled blue symbols) and linear shear modulus ($G$) (filled red symbols) vs. axial strain ($\varepsilon$): 
a) data from measurements on collagen (2.5 mg/ml) ($\circ$) and fibrin (10 mg/ml) ($\bigtriangledown$)\cite{van2016uncoupling}. Collagen stiffens faster than fibrin. The normal stress is set to zero before doing the measurements. b) data from simulations on a 3D phantom lattice for two different $\tilde{\kappa}$ values, $1 \times 10^{-4}$($\circ$) and $1.8 \times 10^{-3}$ ($\bigtriangledown$). c) data from simulations on a 2D phantom lattice for two different $\tilde{\kappa}$ values, $1.1 \times 10^{-4}$ ($\circ$) and $2.2 \times 10^{-3}$ ($\bigtriangledown$). The data in the three panels are normalised by the linear shear modulus of the unloaded network ($\varepsilon=0$), $G_0$. In the simulations, network with lower fibre rigidity which resembles behaviour of collagen, stiffens faster. There is a good agreement between theory and experiment.}
\label{fig:compnopre}
\end{figure}
Similarly, in the simulations, the networks are subjected to successive $1\%$ increments of either compression or extension. After each step, the energy is first minimised and the normal stress $\sigma_N$ is calculated before measuring the linear shear modulus. To measure this, small shear strain ($1\%$) is applied and again the energy of the network is minimised before applying the next axial strain step. 
This process continues over the full $\pm10\%$ axial strain range, with fixed lateral boundary conditions. The results are shown in Fig. \ref{fig:compnopre}. We use  the following sign convention: positive axial strains represent extension, while negative values indicate compression. As expected, when the networks are compressed, both experiments and simulations show positive normal stresses, corresponding to compressive stresses that would lead to expansion in the absence of applied external stress. In Fig. \ref{fig:compnopre}a, normalised shear modulus $G$ and normal stress $\sigma_N$ are shown versus axial strain $\varepsilon$ for both 2.5 mg/ml collagen and 10 mg/ml fibrin networks. As can be seen here, the linear shear modulus changes under axial strain. Both fibrin and collagen samples stiffen under extension but soften under compression\cite{van2016uncoupling}. Moreover, the collagen samples stiffen more rapidly under extension than do the fibrin samples. 
\begin{figure*}[htbp]
\subfloat[]{
 \includegraphics[width=.45\textwidth]{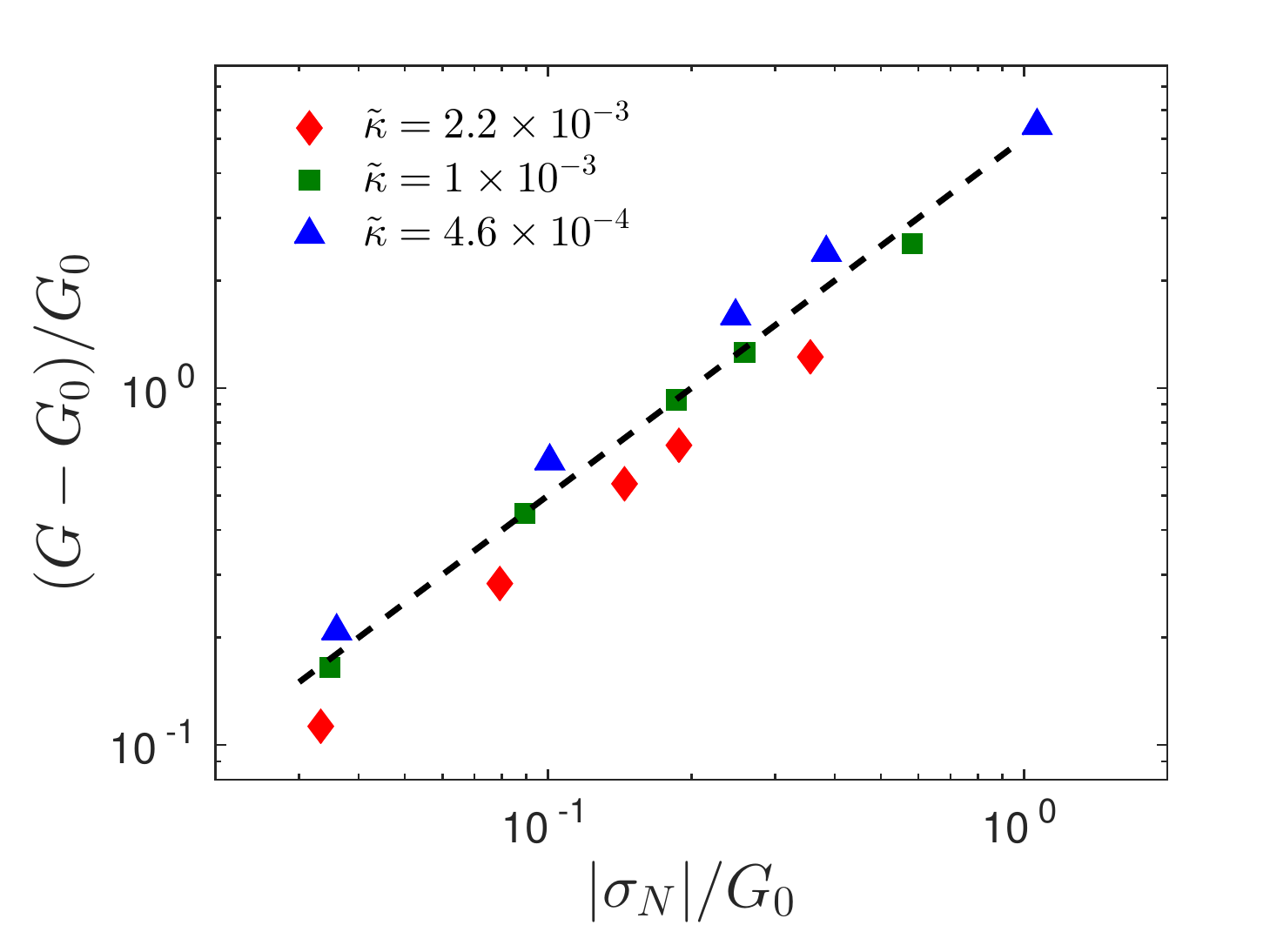}
  \label{fig:simGSigmaN}  
}
\qquad\quad
\subfloat[]{
\includegraphics[width=.45\textwidth]{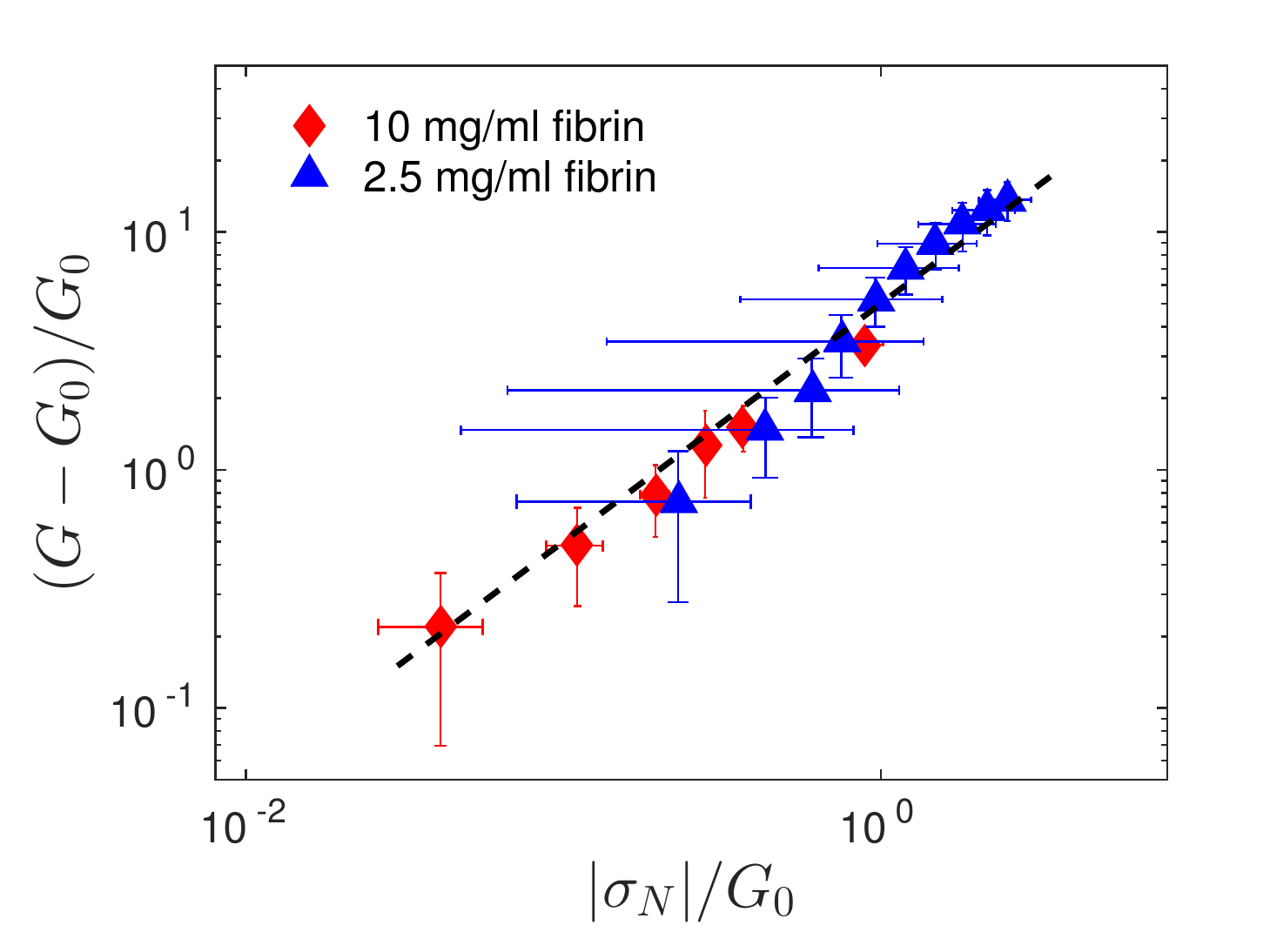}
  \label{fig:expGSigmaN}
}
\caption{(colour online) The normalised increase in linear shear modulus after extension versus the corresponding change in magnitude of normal stress: a) data from simulations on 2D diluted phantom triangular lattice with different $\tilde{\kappa}$. The normalisation is with respect to $G_0$. The dashed line with slope one represents $(G-G_0)=5|\sigma_N|$, corresponding to a good approximation for $\tilde{\kappa}=10^{-3}$. This same line is superimposed in (b) showing excellent quantitative agreement with experiments in that panel. In (b), data are shown for fibrin samples at 2.5 mg/ml and 10 mg/ml. These have also been normalised by $G_0$.}
\label{fig:compGSigmaN}
\end{figure*}
\begin{figure*}[htbp]
\subfloat[]{
\includegraphics[width=.43\textwidth]{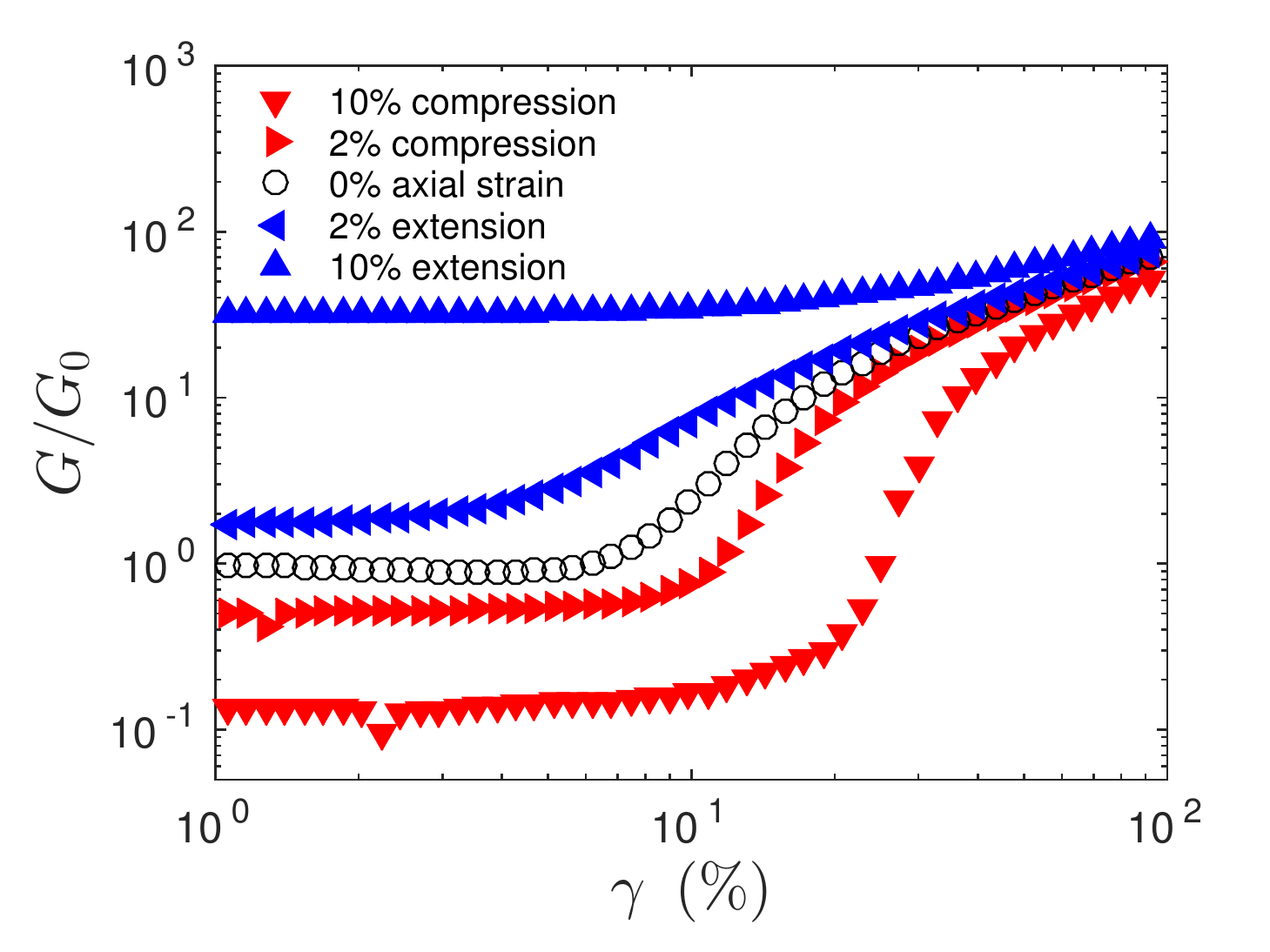} 
  \label{fig:Gvsgammasim}  
}
\qquad\quad
\subfloat[]{
\includegraphics[width=.45\textwidth]{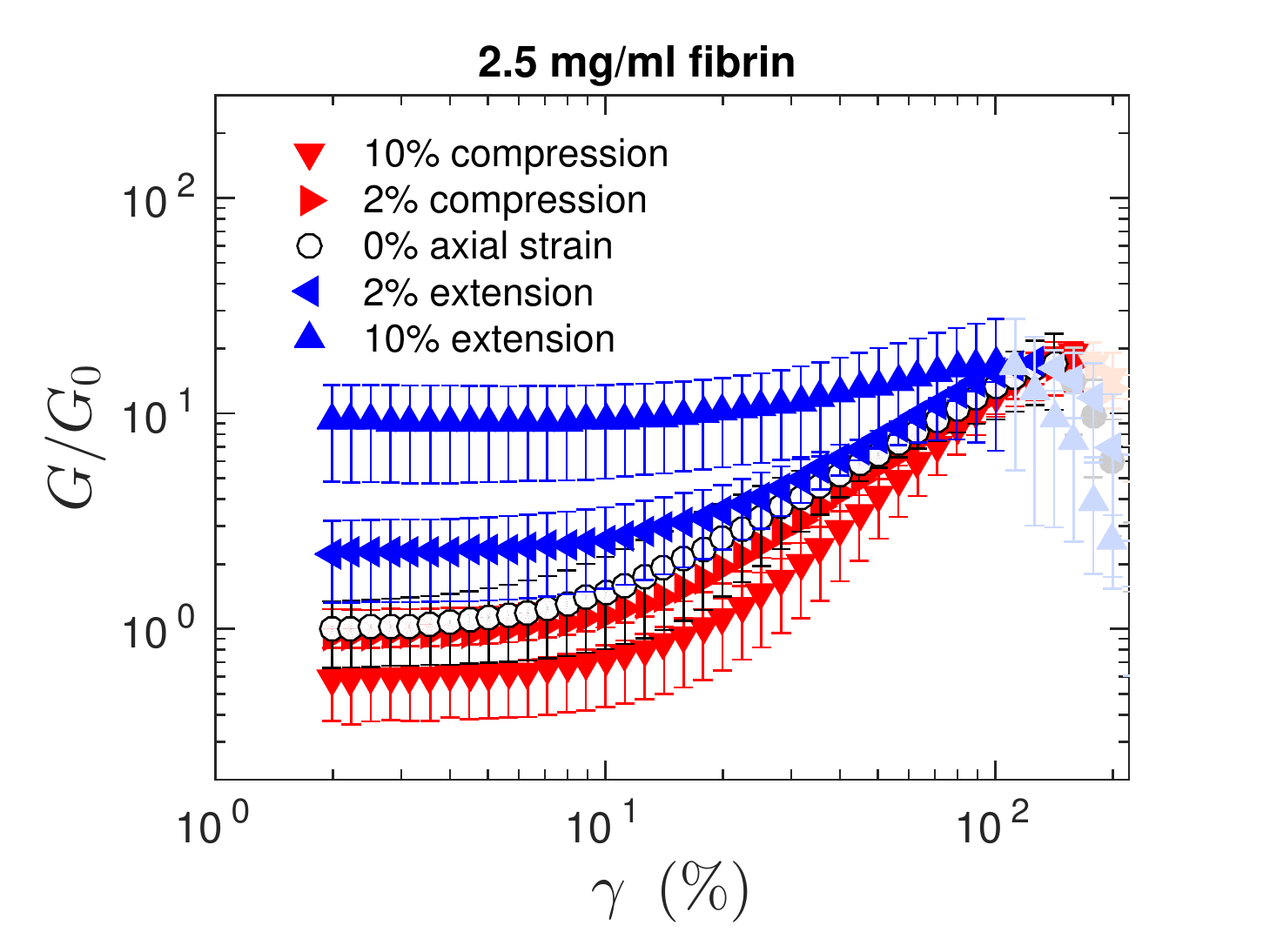}
  \label{fig:Gvsgammaexp}
}
\caption{(colour online) Normalised storage modulus $G$ vs. shear strain $\gamma$ for networks with imposed axial strains. The results for the unloaded networks are shown for better comparison. Linear shear modulus $G_0$ of the unloaded sample is used for normalisation: a) data from simulations on 2D phantom networks with $\tilde{\kappa}=2.2\times 10^{-4}$. b) data from measurements on fibrin (2.5 mg/ml). The error bars are also shown. The downturn in the curves for strains larger than $\sim 100\%$ corresponds to sample detachment from the plates. The corresponding data are shown with lighter colors. Strain-stiffening curves are qualitatively similar for the loaded networks both in experiments and simulation to the unloaded network. The only case which shows a significant deviation is the most extended network $\varepsilon=10\%$ where strain-stiffening is less pronounced. Compressed networks start from lower values thus they are softer. Extended networks lie above the unloaded network and so they are stiffer. The onset of stiffening shifts to the right (higher strains) for the compressed networks and it shifts to lower strains for extended networks with the exception of the most extended sample $\varepsilon=10\%$.}
\label{fig:Gvsgamma}
\end{figure*}%

In the experiments, the normal stress is always set to zero before doing the measurements. This can be seen from the data where we have zero normal stress at $\varepsilon=0$ in Fig. \ref{fig:compnopre}a. In Fig. \ref{fig:compnopre}b, $G$ and $\sigma_N$ for two phantomized 3D FCC networks with fibre rigidities $\tilde{\kappa}=1\times10^{-4}$ and $\tilde{\kappa}=1.8 \times 10^{-3}$ are shown. In Fig. \ref{fig:compnopre}c, $G$ and $\sigma_N$ for two 2D phantom  networks with $\tilde{\kappa}=1.1 \times 10^{-4}$ and $\tilde{\kappa}=2.2 \times 10^{-3}$ are plotted. As noted above, $\tilde{\kappa}$ is expected to be proportional to the volume fraction. Thus, the larger value implied for fibrin is qualitatively consistent with the higher fibrin concentration. In Fig. \ref{fig:compnopre}, the linear shear modulus of unloaded networks is used to normalize the curves. These results are very different from those of tissues\cite{comley2012compressive,fung1967elasticity,lally2004elastic,mihai2015comparison,perepelyuk2016normal,pogoda2014compression}. As can be seen in Fig. \ref{fig:compnopre}, both 2D and 3D networks agree well with the experimental results. Note that the qualitative behaviour of the elastic properties versus axial strain is rather insensitive to concentration in experiments and to fibre rigidity in simulations\cite{van2016uncoupling}. From now on, we show simulation results from 2D networks.
\begin{figure}[htbp]
\centering
\includegraphics[width=.4\textwidth]{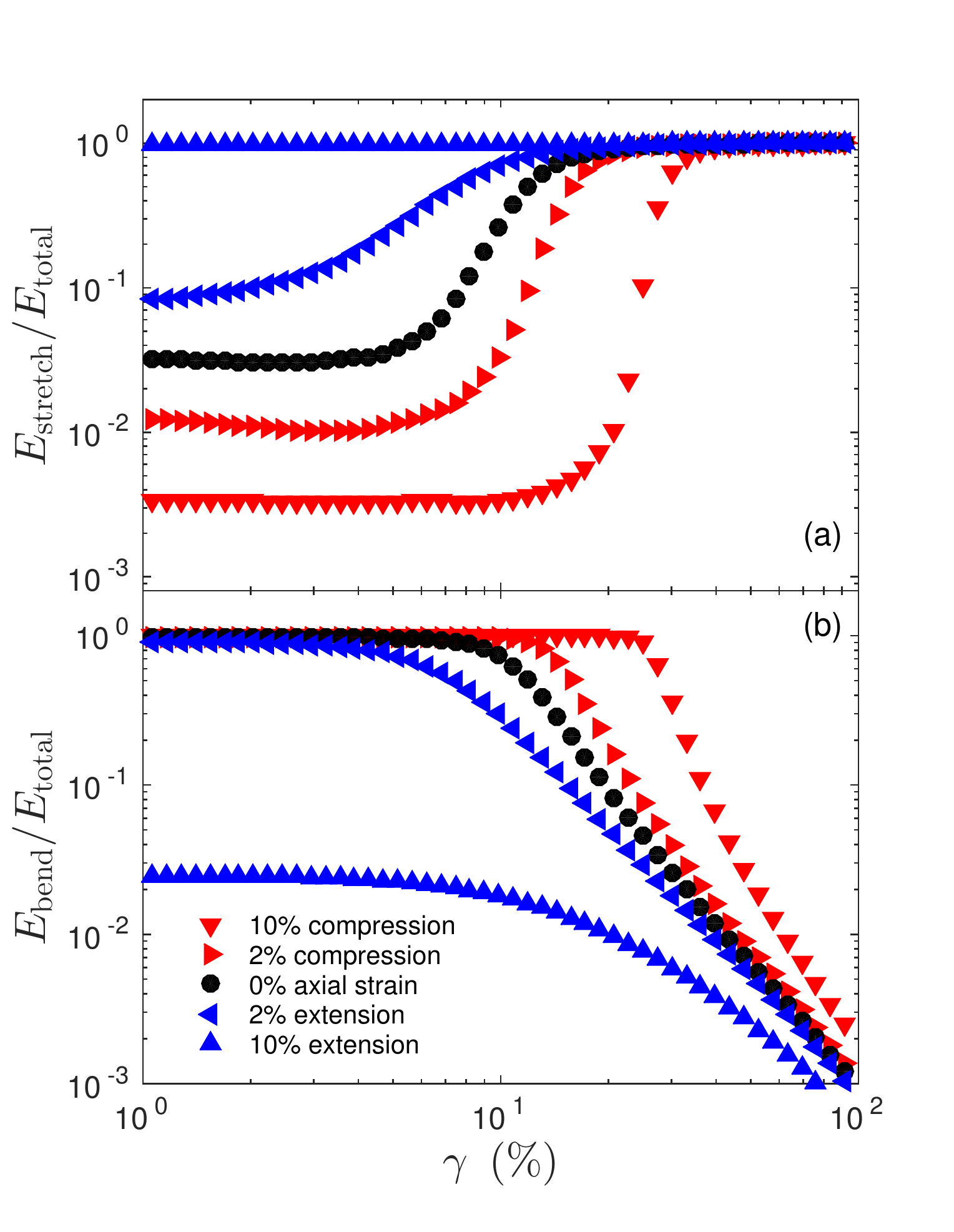}
\caption{
(colour online) Relative energy contributions for a 2D phantom network with $\tilde{\kappa}=2.2\times 10^{-4}$ and different applied axial strain ($\varepsilon$) vs. shear strain, $\gamma$ (a) Relative contribution of total stretching energy to the total elastic energy of the network (b) Relative contribution of total bending energy to the total elastic energy of the network. By applying extensive loads, the stretching energy contribution of the network gets larger and consequently the bending energy contribution gets smaller. In the case of $10\%$ applied axial strain, the network stretching energy is the dominant energy of the network and one sees no transition from bend-dominated to stretch-dominated regimes. Indeed, its response is increasingly becoming affine. The symbols in panel (a) are the same as in panel (b)}
\label{fig:Energyratiovsgamma}
\end{figure}%
It is interesting to note that in Fig.\ \ref{fig:compnopre}, for positive axial strain $\epsilon>0$, the magnitude of the normal stress shows the same trend as the shear modulus: in both the model and experiment, as the shear modulus $G$ increases under extension, so does $|\sigma_N|$, although $\sigma_N$ is negative under (positive) axial strain. In fact, under extension ($\epsilon>0$), the change in the shear modulus $G-G_0$ relative to the unstrained value $G_0$ at $\varepsilon=0$ is predicted to vary approximately linearly with $\sigma_N$, as shown in Fig.\ \ref{fig:simGSigmaN}. 
Here, simulation results are presented for three different values of $\tilde{\kappa}$. The predicted linear dependence of $G-G_0$ on $\sigma_N$ is consistent with the 
experimental results of Fig.\ \ref{fig:compnopre}. In Fig.\ \ref{fig:expGSigmaN}, the experimental $G-G_0$ is plotted versus $|\sigma_N|$ for 10 mg/ml and 2.5 mg/ml fibrin gels. The experimental data are normalised by the square of their concentration $c^2$. Prior experiments have shown an approximate quadratic dependence of shear modulus on concentration for collagen\cite{piechocka2011rheology,motte2013strain} and fibrin\cite{piechocka2010structural} $G \sim c^2$. 
Interestingly, in addition to the consistency with the predicted linear scaling of $G-G_0$ with $\sigma_N$, we also note the good experimental agreement with the predicted prefactor, $(G-G_0)\simeq 5|\sigma_N|$ for $\tilde{\kappa}=10^{-3}$.
Prior work on collagen\cite{Licup04082015} and model networks with compliant crosslinks\cite{heidemann2015elasticity} have also reported an approximate linear scaling of modulus with the normal stress.

\begin{figure*}[htbp]
\subfloat[]{
\includegraphics[width=.43\textwidth]{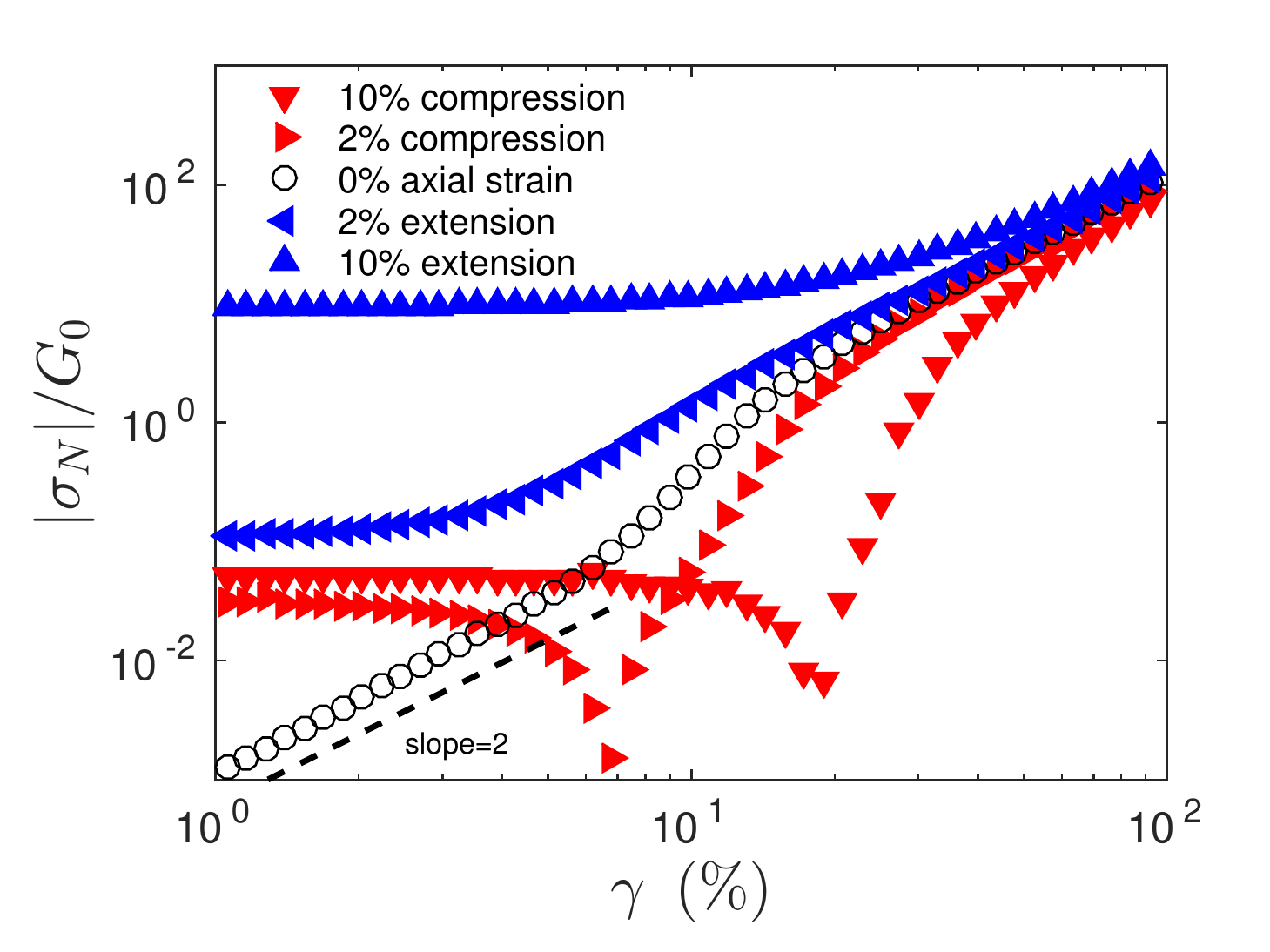} 
  \label{fig:sigmanvsgammasim}  
}
\qquad\quad
\subfloat[]{
\includegraphics[width=.45\textwidth]{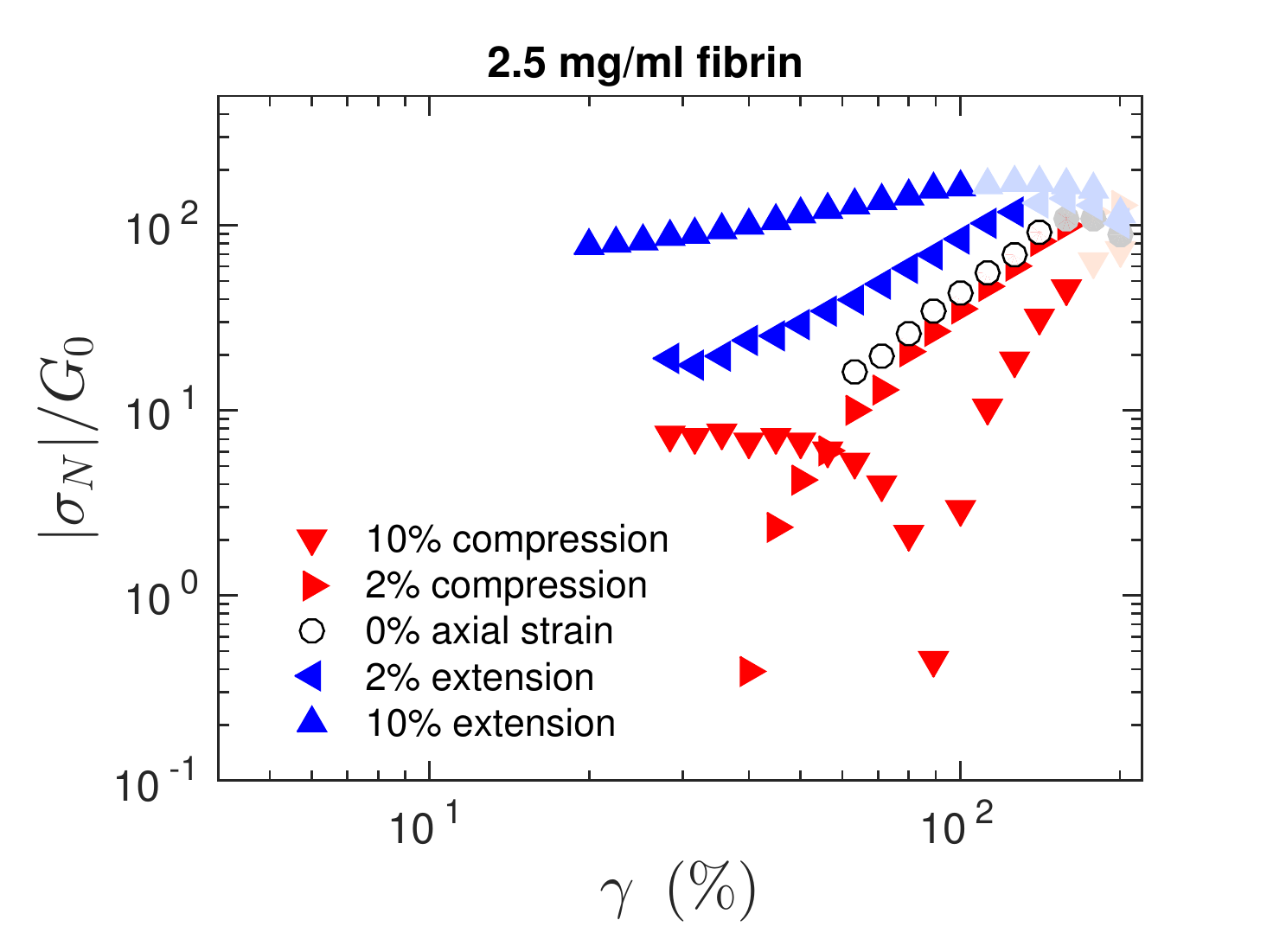}
  \label{fig:sigmanvsgammaexp}
}
\caption{(colour online) Absolute value of the normal stress ($|\sigma_{N}|$) vs. shear strain $\gamma$ for different applied axial strain a) data from simulations on a 2D phantom network with $\tilde{\kappa}=2.2\times 10^{-4}$. We can see that the normal stresses in the cases with imposed uniaxial compression and extension do not show the expected $\gamma^2$ dependence in the no axial case. b) data from measurements on fibrin (2.5 mg/ml). The data corresponding to the downturn in storage modulus curves (see Fig. \ref{fig:Gvsgammaexp}) are shown with lighter colors. In both theory and experiment, the normal stresses in the extended networks are always negative but the normal stresses in the compressed networks change sign from the initial positive to negative values. The dip in the compressed networks show the sign change of the normal stress. The normal stress values are normalised by the linear shear modulus $G_0$ of the network with no imposed axial strain.}
\label{fig:normalstressvsgamma}
\end{figure*}%

\subsection{Strain-stiffening and its dependence on prestress}\label{B}
We now consider the following questions: (1) How do the shear strain stiffening curves change with applied extension and compression? (2) How does the onset of shear strain stiffening change for different prestressed states? (3) How does the normal stress vary when we apply nonlinear shear deformation to prestressed networks? The samples are first subjected to an applied static compression/extension in a strain-controlled rheometer with parallel plates. The axial strain is applied by changing the gap size between the two plates. As in our previous measurements, volume change is allowed by surrounding the sample with solvent. An oscillatory shear strain of constant frequency of $1 \enskip \mathrm{rad/s}$ with an increasing magnitude is subsequently applied. The shear storage modulus is then measured. In the simulations, 2D and 3D diluted phantom networks are first compressed/extended with similar amounts of axial strains as in the experiments, after which the energy of the network is minimised. Normal stresses are calculated and then by keeping the axial strain fixed, the network is subjected to increasing shear deformation from $1\%$ until $100\%$ in logarithmic steps. After each shear step, the energy of the system is again minimised and the storage modulus is calculated.

In Fig. \ref{fig:Gvsgammasim}, the strain-stiffening curves from a 2D diluted phantom lattice with fibre rigidity $\tilde{\kappa}=2.2 \times 10^{-4}$ are shown for different amounts of axial strains. In Fig. \ref{fig:Gvsgammaexp}, the nonlinear strain-stiffening curves from 2.5 mg/ml fibrin samples are presented for different amount of prestress. The stiffening curve for $0\%$ axial strain is also shown in both panels for comparison. Both simulation and experimental results are normalised by the linear shear modulus $G_0$ of the unloaded network. As can be seen in Fig. \ref{fig:Gvsgamma}, the experimental results show good qualitative agreement with the simulation results. Extended networks are stiffer and compressed networks are softer. The variation in the onset of stiffening for different axial strains also shows similar behaviour as in the simulation results. By applying compression the onset of stiffening shifts towards larger strains relative to the case without any axial load suggesting that nonlinear behaviour is delayed. The more compressed the sample, the larger the shear strains at which the nonlinear behaviour is seen. This can be understood in the simple physical picture that when the samples are compressed, the fibres in the network buckle or bend due to smaller energy cost for bending than compression (see Fig. \ref{fig:Energyratiovsgamma}a and \ref{fig:Energyratiovsgamma}b). Due to bending, the end to end distance of the fibres become smaller than the contour length which results in an excess length\cite{Licup04082015,licup2016elastic}. The excess length is related to the onset of stiffening. Larger excess length leads to larger strains for the onset of stiffening. Thus, compression shifts the onset of stiffening to larger shear strains. In contrast, the inverse happens for extended networks. The smaller excess length results in smaller onset of stiffening. The shift in the onset to lower strains with the applied extension occurs only over a limited range of axial extension. Beyond a certain extension, it appears that the strain threshold for the onset of stiffening increases with the applied extension. This happens because beyond a certain extension, no excess lengths can build up in the fibres. In fact, after a sufficiently large extension, the shear response of a network can be captured by the affine prediction. In an affinely deformed network, the elastic response is only governed by stretching modes for any applied shear strain. In this case, the onset of stiffening is determined by geometric alignment of fibres, which is attained at large shear strains. The same effect was observed in Ref.\cite{van2016uncoupling}. Extracellular networks of collagen and fibrin show different onsets under axial extension and the reason lies in the larger extension imposed on the fibrin samples compared to collagen.

It is also informative to look into the variation of the normal stresses during shear stiffening and compare the results under varying amounts of prestress. Different studies investigated the normal stresses of biopolymer networks when sheared. The normal stresses of these networks are negative under shear which are opposite in sign (direction) compared to those measured from most elastic solids. This is known as the Poynting effect\cite{poynting1909pressure,poynting1912changes}. From symmetry arguments, the normal stresses (if analytical) should only be functions of even powers of $\gamma$. For low strains, $\sigma_N \sim \gamma^2$ is expected based on symmetry considerations\cite{janmey2007negative,kang2009nonlinear,heussinger2007role,conti2009cross}. The absolute values of normal stresses of the same networks as in Figs. \ref{fig:Gvsgamma} and \ref{fig:Energyratiovsgamma} are shown in Fig. \ref{fig:normalstressvsgamma}. Again, for comparison, the results of the network with no axial load is also shown. As expected, the normal stresses from simulation results for networks without axial load initially show the $\gamma^2$ regime (see Fig.\ref{fig:sigmanvsgammasim}). Extension and compression loads introduce opposite effects on the axial response of the network. When networks are extended, they tend to pull the boundary downward (negative normal stress) while compression induces an upward (positive normal stress) response. The normal stress of the compressed networks start from positive values while the extended ones show negative values for low shear deformations. With increasing shear strain, the extended networks show even larger negative normal stresses while the initial positive normal stresses of the compressed networks decrease in magnitude, then cross over at zero to switch sign. The dip in the normal stress of the compressed networks (in absolute value) show the sign change.  It might seem that the strain at which the sign change of the normal stress occurs coincides with the onset of stiffening. However we find that it is not the case. The experimental results show good agreement with simulation (see Fig. \ref{fig:sigmanvsgammaexp}), although the normal stresses at low strain values were difficult to resolve due to device limitations. The data corresponding to the downturn in storage modulus curves of Fig. \ref{fig:Gvsgammaexp} are shown with lighter colors.

\begin{figure}[t]
\includegraphics[width=.4\textwidth] {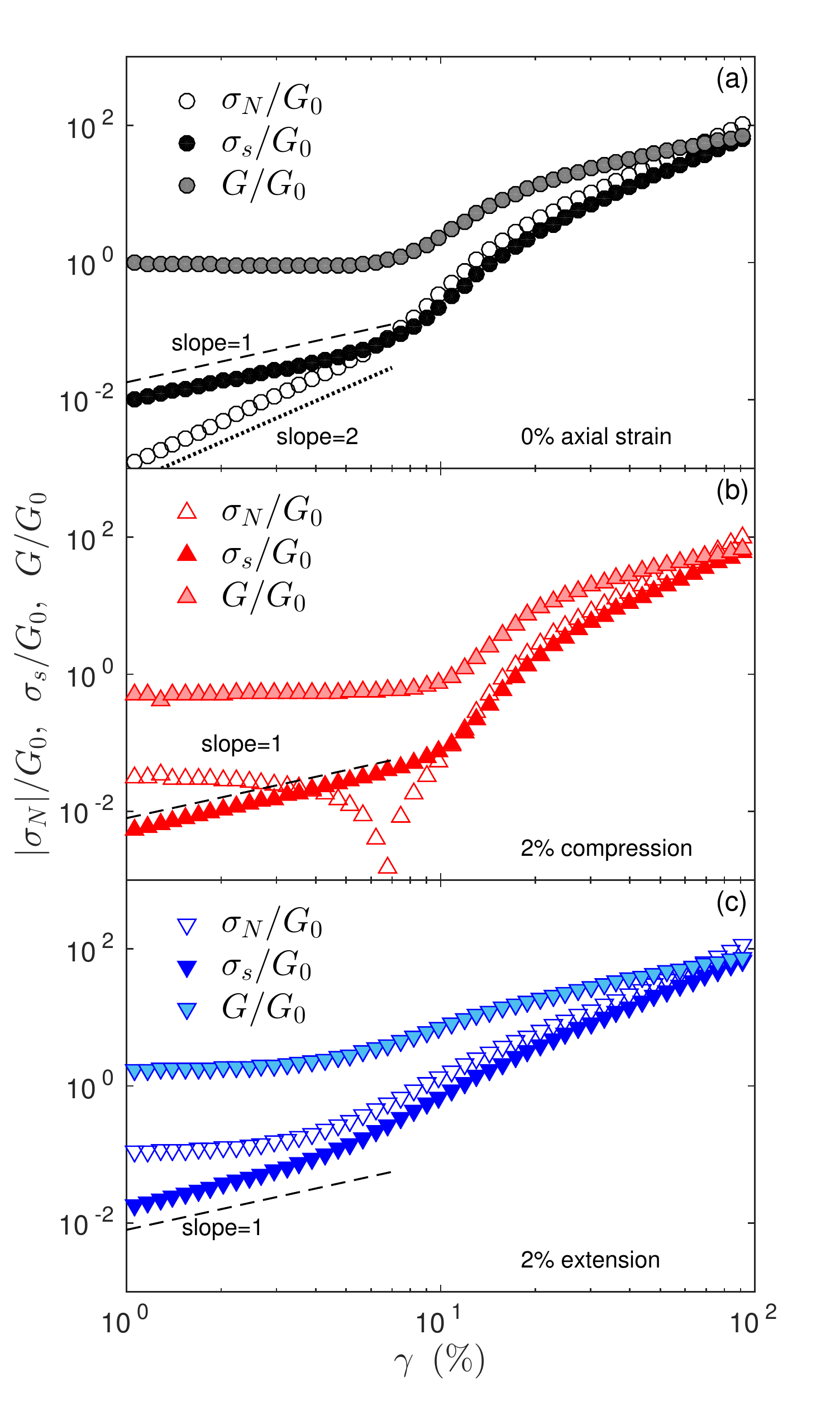}
\caption{Comparison of the normalised shear $\sigma_s$ and normal stresses $\sigma_N$ and storage modulus $G$ versus shear strain $\gamma$. Linear shear modulus $G_0$ of the unloaded network has been used for all the normalisations. At the onset of stiffening, normal and shear stresses become comparable for unloaded networks which has been noticed both theoretically and experimentally\cite{janmey2007negative,Licup04082015}. This is still the case for both unloaded and compressed networks (panels (a) and (b)). For extended networks, this is not the case (panel (c)).}
\label{fig:onsetcomp}
\end{figure}

It has been demonstrated that for the network without an imposed axial strain, shear and normal stresses become comparable at the onset\cite{janmey2007negative,Licup04082015}. In Fig. \ref{fig:onsetcomp}, we show shear stress, storage modulus and normal stress versus shear strain. We consider the three cases: no axial strain, compression and extension. The linear shear modulus $G_0$ has been used for the normalisation. Normal and shear stresses become comparable at the onset of stiffening. This holds for unloaded and compressed networks (see Figs. \ref{fig:onsetcomp}a and \ref{fig:onsetcomp}b). However, for extended networks, this is not the case (see Fig. \ref{fig:onsetcomp}c). Here, at the onset of stiffening, the shear stress is still smaller than the normal stress. It is important to note that the onset of stiffening does not coincide with the shear strain at which normal stress changes sign as seen in Fig. \ref{fig:onsetcomp}b.

\begin{figure}[t!]
\centering
\includegraphics[width=0.4\textwidth]{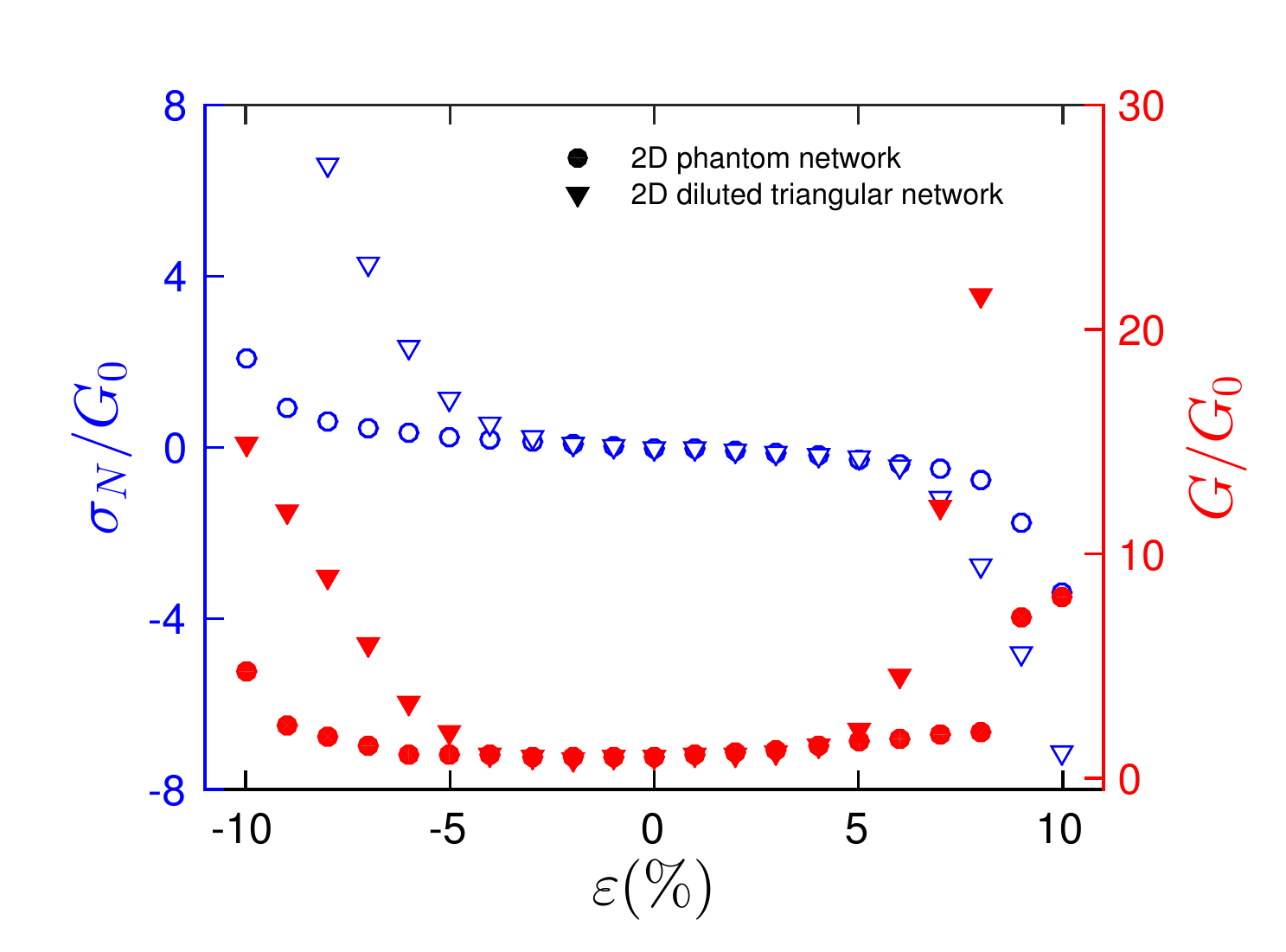}
\caption{(colour online) Normalised normal stress (unfilled blue circles) and shear modulus (filled red circles) of a 2D phantom triangular network with fibre rigidity, $\tilde{\kappa}=10^{-4}$ under uniaxial extension and compression with global volume constraint boundary condition. The data are normalised by the shear modulus at zero axial strain $G_0$. We see stiffening for both compression and extension. Here we use networks of $30^2$ nodes.}
\label{fig:GVC}
\end{figure}%

\subsection{Tissues and global volume constraint boundary condition}\label{C}
As seen in the previous subsections, experimental and computational results show softening for compression and stiffening for extension of biopolymer networks in solvent, which is in contrast with experimental reports of stiffening under compression for tissues\cite{mihai2015comparison,perepelyuk2016normal,pogoda2014compression}. This property can be seen in our networks if we apply appropriate boundary conditions for incompressibility, in which the sample expands (contracts) laterally under axial compression (expansion). For incompressible 2D networks, the lateral strain is equal and opposite to the axial strain, corresponding to a Poisson ratio of unity. 

In Fig. \ref{fig:GVC}, we have used the 2D network model similar to the previous subsections with the only difference being the global volume constraint, to impose the incompressibility condition. The difference between the two curves is their corresponding network structure. We show the result for a 2D phantom network with $\langle z \rangle=3.2$ and a 2D diluted triangular network with $\langle z \rangle=3.3$ for which this connectivity is reached by random bond removal. We observe stiffening for both axial compression and extension. The strain at which stiffening starts (about $5\%$ for 2D phantom network and about $8\%$ for 2D diluted triangular network) or the shape (steepness) of the curve is dependent on the network structure as well as fibre rigidity. In 3D networks, considering the global volume constraint, the lateral strain is not the same as axial strain. Despite the Poisson ratio of one half in 3D, one would expect stiffening for both compression and extension. 

\section{Conclusions}\label{Conclusions}
We have studied the elastic properties of networks to which axial strain has been applied. Specifically, we studied, both experimentally and in simulation, normal and shear stresses, as well as strain-stiffening and the linear shear modulus. The experimental results from reconstituted networks of fibrin and collagen have been compared with results from lattice-based networks with physiological connectivity, in both 2D and 3D. Networks in both 2D and 3D give similar behaviour for the same connectivity. In both cases, we find good qualitative agreement with experiments. In the experiments, the rheometer is surrounded with buffer allowing for water to freely move in or out. In simulations, fixed boundary conditions are used to be consistent with experiments. By using fixed boundary conditions, applied extension or compression results in a volume change. Prestress resulting from the applied extension and compression, strongly affects the network elastic response. Softening due to compression and stiffening due to extension are observed for both experiments and simulations. By applying a global volume constraint in order to account for the volume-preserving aspect of tissue\cite{fung1967elasticity}, our simulation results show stiffening for both extension and compression. The linear shear modulus increases with the normal stress and exhibits an approximately linear scaling with normal stress both in experiment and in simulation. 

The strong dependence of the mechanics of extracellular networks on prestress can be expected to have important consequences for both fundamental tissue mechanics, as well as for tissue engineering. The softening of compressed samples and the dependence of the strain onset of stiffening, for instance, are likely to be important mechanical parameters for synthetic tissue scaffolds. Network simulations are powerful techniques to gain more insight into these mechanical parameters for the design of such scaffolds and other biocompatible materials.



\section*{ACKNOWLEDGMENTS}  M.V., A.S. and A.J.L. were supported by Stichting voor Fundamenteel Onderzoek der Materie which is part of the Nederlandse Organisatie voor Wetenschappelijk Onderzoek. A.S. was also partly supported by NanoNextNL. PAJ and AvO were supported by grants  NIH EB017753 and NSF-DMR-1120901. AvO was supported by a Fulbright Science and Technology Award, and the Prins Bernhard Cultuurfonds-Kuitse Fonds. The simulations were done on the Dutch national e-infrastructure with the support of SURF Cooperative. 

\bibliography{FollowupPaul} 
\bibliographystyle{rsc} 

\end{document}